# Bandwidth Modeling and Estimation in Peer to Peer Networks

Kiarash Mizanian, Mehdi Vasef[*] and Morteza Analoui

School of Computer Engineering

Iran University of Science and Technology, Tehran, IRAN

*Student Member, IEEE

mehdi.vasef@ieee.org

*Abstract*-Recent studies have shown that the majority of today's internet traffic is related to Peer to Peer (P2P) traffic. The study of bandwidth in P2P networks is very important. Because it helps us in more efficient capacity planning and QoS provisioning when we would like to design a large scale computer networks. In this paper motivated by the behavior of peers (sources or seeds) that is modeled by Ornstein Uhlenbeck (OU) process, we propose a model for bandwidth in P2P networks. This model is represented with a stochastic integral. We also model the bandwidth when we have multiple downloads or uploads. The autocovariance structure of bandwidth in either case is studied and the statistical parameters such as mean, variance and autocovariance are obtained. We then study the queue length behavior of the bandwidth model. The methods for generating synthetic bandwidth process and estimation of the bandwidth parameters using maximum likehood estimation are presented.

## 1. Introduction

The term "peer-to-peer" refers to a class of systems and applications that employ distributed resources to perform a function in a decentralized manner. Benefits of peer to peer systems are cost sharing/reduction, resource aggregation, increased autonomy, anonymity/privacy of the users and finally enabling ad hoc communication and collaboration [1].

Napster [2] was the first popular peer to peer service. This service has allowed hundreds of thousands of users to efficiently share MP3 formatted files. The success of Napster was a big motivation and several other peer to peer file sharing systems were introduced. These include KaZaA [3], Gnutella [4], and eDonkey [5] and BitTorrent [6].

According to Cache Logic [7] by the end of 2004, BitTorrent accounted for as much as 30% of all Internet traffic. Peer to peer represented 60% of Internet traffic at the end of 2004. Most dominant P2P

---

[*] Mehdi Vasef is currently a graduate student at University of Duisburg-Essen, Duisburg, Germany

systems are BitTorrent, eDonkey and Gnutella. The number of people that use peer to peer file sharing is growing. In January 2005, 2,975,477 online eDonkeyy2k users were reported. In January 2006, the number was increased to 3,351,754 [8].

The number of P2P users , the average file size transported in P2P file sharing systems and percentage of overall network traffic include P2P network traffic are growing, so bandwidth management plays an inevitable role in designing efficient computer networks nowadays. We still do not have accurate models for P2P bandwidth. In this paper, we propose a novel model for modeling P2P bandwidth. To achieve this goal, we should consider both the customers and the share of these customers. The customers are peers and the shares are their traffics in a given network. The second part was accomplished in the authors' previous work. So our starting points deeming the behavior of peers and pondering it. We use OU type process to describe the peer behavior. We then proceed to model the P2P bandwidth. We use the stochastic calculus approach in our proposed model. Some statistical parameters of the bandwidth model are derived. We also present a model for the total bandwidth .It is shown that the total bandwidth asymptotically has a LRD property.  We also derive the length of the buffer fed by the bandwidth. A method for generating synthetic bandwidth is presented. Estimation of bandwidth parameters is another contribution of this paper.

More accurately the contributions of this paper are:

- Modeling the P2P bandwidth and obtaining some parameters such as mean and variance
- Modeling total P2P bandwidth and obtaining some parameters such as mean and variance
- Study of autocovariance in bandwidth model both in individual and total case and
 Showing that the total bandwidth is asymptotically a Long Range Dependent process
- Obtaining Length of queue fed by a bandwidth process
- Synthetic generation of bandwidth process
- Estimating bandwidth model parameters using MLE method

The rest of paper is organized as follows. In section 2, we review related works. In section 3, we introduce the mathematics required to understand the model. In section 4, we explain our proposed model. Queuing analysis of the proposed model, synthetic bandwidth generation and the estimation of bandwidth parameters are also studied in section 4. Finally section 5 concludes the paper.

## 2. Related Works

To best of our knowledge the mathematical model for P2P bandwidth is an unexplored area in the literature. The next paragraphs illustrate the related works that have been done in P2P traffic modeling, modeling the behavior of peers and their impact on the P2P system properties.

[9] proposes a stochastic differential equation approach for modeling the behavior of Peers. The steady state behavior of the peers is investigated. But the bandwidth model is not studied. [10] Proposes a fluid model that characterizes the number of peers in BitTorrent like networks. The arrival and departure processes for sinks and sources are Poisson processes. Another assumption is that all the peers have the same uploading/sinking bandwidth. [11] Proposes a fluid model for BitTorrent system but in comparison to [10], the peer arrival rate is exponential. Their finding is that the existing BitTorrent system provides

poor service availability, fluctuating sinking performance, and unfair services to peers. Their model has revealed that these problems are due to the exponentially decreasing peer arrival rate. A stochastic fluid model [12] is proposed to study performance of peer to peer web cache (SQIRREL) and cache cluster by extending [10]. The [13, 14] use Markovian model to describe P2P file sharing system .But they don't work well in some P2P system such as BitTorrent. In addition the steady state behavior of peers is not studied.

Recently [15] have proposed a model for P2P traffic. The model is based on Alternating Fractal Renewal Process (AFRP) such that the ON/OFF periods are kind of power law distribution. The sink/source rates are considered constant. The proposed model can capture long range dependent as well as heavy tailedness property.

## 3. Mathematic Preliminaries

### 3.1. Long Range Dependence

The self similar nature of network traffic was discovered by Will Leland [16]. Mark Crovella Showed that World Wide Web exposed self similar property [17]. Afterwards plenty of papers have been published on the impact of LRD on traffic modeling and queuing performance in computer networks. Here we present some mathematical definitions of LRD based on auto-covariance function.

Let $\psi(k)$ denotes auto-covariance function in a process with LRD property, that is $1/2 \leq H \leq 1$
Then we have $\psi(k) \sim ck^{2(H-1)}$ [18]. The H constant is called self similarity parameter or Hurst parameter.

Suppose we have an aggregated process $X^m$ of $X$ at aggregation level m, then we have $\psi_t = \sum_{k=1}^{\infty} \psi(k)$
If $\psi(k) \to \infty$, then $X$ is has a LRD property [18].

### 3.2. Ornstein Uhlenbeck Process

The Ornstein Uhlenbeck process (or mean reverting process) is a stochastic process defined by the following stochastic differential equation [19, 21]:

$$dS_t = \gamma(\mu - S_t)dt + \sigma dW_t \tag{1}$$

$$S_t(0) = S_0 \tag{2}$$

such that:

$\gamma: mean\ reversion\ rate$

$\mu: mean$

$\delta$: volatility

$\{W(t)|t \geq 0\}(t)$ is a Wiener process

$$E(s) = E(s_0) \tag{3}$$

$$Var(s) = \frac{\sigma^2}{2\gamma} \quad \text{are long term mean and variance} \tag{4}$$

The covariance function of OU process is given by: $cov(S_{t_1}, S_{t_2}) = \frac{\sigma^2}{2\gamma} e^{-\gamma(t_1+t_2)}$ (5)

The OU process can be simulated (generated synthetically) by the following formula:

$$S(t) = S(0)e^{-\gamma t} + \mu(1 - e^{-\gamma t}) + \sigma \sqrt{\frac{1-e^{-2\gamma t}}{2\gamma}} N(0,1) \tag{6}$$

$S(t)$: OU process at future time t

$S(t)$: current value of OU process

$N(0,1)$: random samples of a normal distribution

### 3.3. Stochastic Calculus

In stochastic calculus the methods of calculus such as integration and differentiation is extended to stochastic processes such as Wiener process. There are two approaches: Ito calculus and Stratonovich calculus. We use Ito calculus in this paper [20]:

$$Y_t = \int_0^t H_s \, dX_s \tag{7}$$

$X$: Brownian motion (Semi − Martingale)

$H_s$: Stochastic process

It can be defined in a way like Rieman-Stieltjes integral:

$$\int_0^t H \, dB = \lim_{n \to \infty} \sum_{t_{i-1}, t_i \in \pi_n} H_{t_{i-1}}(B_{t_i} - B_{t_{i-1}}) \tag{8}$$

The limit can be shown that converges in probability. The proof is out of the scope of this paper. The interested reader is referred to [20] for more information.

## 4. Bandwidth Modeling in P2P networks

### 4.1 Bandwidth Model for P2P Networks

Results of studies demonstrate that the evolution of peers is very similar to the OU process [10]. They tend to approach to a definitive value in steady state. They are very similar to stock returns that are modeled with OU process. This leads us to choose the OU process as a model for evolution of peers. We consider some assumptions in proposing the model for P2P bandwidth modeling. These assumptions can be summarized as follows:

- The evolution of peers is modeled with OU process
- The peers download or upload rate is constant
- We use the P2P traffic model for an individual peer level proposed in our previous work [15]
- The behavior of peers is independent of their individual traffic

First we model the bandwidth for individual peer. We have the following definition for individual bandwidth in P2P networks.

*Definition1*: Assume we would like to perform a single operation on a peer in predefined set of peers. This operation may be downloading or uploading some file by a peer. The bandwidth in this peer's link and the involved peers is called individual bandwidth. We represent individual bandwidth with $Bw$.

In order to model bandwidth we need two elements. The customers (or peers) and the share of them (the corresponding P2P traffic). With reference to previous section, we mentioned that OU process was used for modeling behavior of peers. The only item remains is the share of these peers (P2P traffic). In [15] we propose a model for P2P traffic in the case the traffic is heavy tail and possesses a LRD property. We suppose that we deal with such traffic. So regarding these elements, we propose the P2P bandwidth that is defined as a stochastic integration of traffic with respect to OU process:

$$Bw = \sum B(t_s) \Delta_s \tag{9}$$

$$\tau - t' = t_s \tag{10}$$

$$Bw = \sum B(\tau - t') \Delta_s \tag{11}$$

In the general case the summation changes into integral: $Bw = \int_{-\infty}^{t} B(\tau - t') \, dS(s)$ \hfill (12)

(12)

$$dS(s) = \lim_{r \to s} S(r, s) \tag{13}$$

$S$ is defined in infinitesimal ranges. $B$ is the share of every peer or simply the P2P traffic.

Let's scrutinize the definition. Before delving into the formula, we rewrite the inner parentheses of $B$:

$$\tau - t' = \tau - (t + t_2) = (\tau - t_2) - t = \tau' - t \tag{14}$$

So $\quad \tau - t' = \tau' - t \tag{15}$

$$Bw = \int_{-\infty}^{t} B(\tau' - t)\, dS(s) \tag{16}$$

Using definition of OU process and substituting with $dS(s)$, we have:

$$Bw = \int_{-\infty}^{t} B(\tau' - t)\, (\gamma(\mu - S_t)dt + \sigma dW_t) \tag{17}$$

$$Bw = \int_{-\infty}^{t} B(\tau' - t)\, \gamma\mu\, dt - \int_{-\infty}^{t} B(\tau' - t)\gamma S_t\, dt + \int_{-\infty}^{t} B(\tau' - t)\sigma \tag{18}$$

We suppose that the OU process is zero mean. So the first term of Eq.18 is omitted.

$$Bw = -\gamma \int_{-\infty}^{t} B(\tau' - t) S_t\, dt + \sigma \int_{-\infty}^{t} B(\tau' - t)\, dW_t \tag{19}$$

Comparing the first integral with the convolution of two functions that is represented by:

$$f * g = \int_{-\infty}^{\infty} f(\tau) g(t - \tau)\, d\tau \tag{20}$$

We have:

$$Bw = -\gamma\big(B(\tau' - t) * S(t)\big) + \sigma \int_{-\infty}^{t} B(\tau' - t)\, dW_t \tag{21}$$

$$w = -\gamma\big(B(\tau' - t) * S(t)\big) - \sigma \int_{-\infty}^{t} B(p)\, dW_p \tag{22}$$

$$Bw = -\gamma\big(B(\tau' - t) * S(t)\big) - \sigma \lim_{n \to \infty} \sum_{p_{i-1}, p_i \in \pi_n} B_{p_{i-1}}(W_{p_i} - W_{p_{i-1}}) \tag{23}$$

Note that since $Bw$ function range is positive, we consider absolute values or mathematically:
$Bw = |Bw| \tag{24}$

Finally referring to section 3, the second integral can be computed like Reiman-Stieltjes integral:

$$Bw = \gamma\big(B(\tau' - t) * S(t)\big) + \sigma\lim_{n \to \infty} \sum_{p_{i-1}, p_i \in \pi_n} B_{p_{i-1}}(W_{p_i} - W_{p_{i-1}}) \tag{25}$$

It can be easily proved that the limit converges in probability and the interested reader is referred to [20] in stochastic integrals. Considering Eq.25, $Bw$ is a parsimonious model. Only some parameters are needed to describe $Bw$. Now we derive some statistical properties of the bandwidth process such as mean, variance and autocovariance function.

**Lemma1**: The mean and variance of bandwidth process are represented by:

$$E(Bw) = \gamma(E(B) + E(S)) + \sigma K' \qquad (26)$$

$$Var(Bw) = \gamma(Var(B) + Var(S)) + \sigma K' \qquad (27)$$

**Lemma2**: The autocovariance of bandwidth process is represented by

$$ACV(Bw) \quad C_1 k^{2\left(\frac{4-\min(n_0,n_1)}{2}+\varepsilon-1\right)}$$

and the bandwidth process has LRD property.

### 4.2 Superposition of Bandwidth Processes

Having modeled individual bandwidth in P2P networks, we would like to study aggregated bandwidth in P2P networks. The aggregated bandwidth process is defined below:

*Definition 2*: Assume we would like to perform multiple operations on a peer in predefined set of peers. These operations may be downloading or uploading multiple files by a peer. The bandwidth in this peer's link and the involved peers in these operations is called aggregated bandwidth. We represent the aggregated bandwidth with $Bw_t$.

In this section we model aggregated bandwidth and derive some properties such as mean, variance and autocovariance function. According to Definition2, we have: $Bw_t = \sum Bw$ $\qquad (28)$

The following lemma demonstrates how one can obtain the mean, variance of the aggregated bandwidth process.

**Lemma3**: The mean and variance of aggregated bandwidth process are represented by:

$$E(Bw_t) = \sum_{i=1}^{N}\left(\gamma_i(E(B_i) + E(S_i)) + \sigma_i K'_i\right) \qquad (29)$$

$$Var(Bw_t) = \sum_{i=1}^{N}\gamma(Var(B) + Var(S)) + \sigma K' \qquad (30)$$

Recall from section 3 that: $\psi_t = \sum_{k=1}^{\infty}\psi(k)$
(31)

$$ACV_t(Bw) \quad \sum_{i=1}^{\infty} C_1 k_i^{2\left(\frac{4-\min(n_0,n_1)}{2}+\varepsilon-1\right)} \tag{32}$$

Since the individual bandwidth processes have LRD property, this sum is asymptotically infinite. So the auto-covariance function of aggregated bandwidth process is asymptotically infinite.

The following lemma explains the queuing behavior of bandwidth process.

**Lemma4**: The queue length of bandwidth process is given by:

$$P(V(t) \succ x) \quad \exp\left(-\left[\frac{-1}{2m\left(\lambda(Var(B)+Var(B)+\sigma K')\right)\left(\frac{\min(n_0,n_1)}{2}-1-\varepsilon\right)^2}\left(\frac{(1-m)\left(\frac{\min(n_0,n_1)}{2}-1-\varepsilon\right)}{\frac{4-\min(n_0,n_1)}{2}+\varepsilon}\right)^{(4-\min(n_0,n_1)+2\varepsilon)}\right]x^{(\min(n_0,n_1)-2-2\varepsilon)}\right) \tag{33}$$

### 4.3 Bandwidth Modeling in Multi-service P2P Networks

Suppose we have a multiservice P2P network- the P2P network with audio, video, image and etc. In this case we define vector valued process $Bw$ for bandwidth modeling in multiservice P2P networks.

$$Bw = (Bw_1 \quad Bw_2 \quad Bw_3 \ldots \quad Bw_n)$$

$$Bw = \int_{-\infty}^{t} B(\tau - t') dS(s) \tag{34}$$

such that elements of this vector is defined as Eq.12.

$$B = (B_1 \quad B_2 \quad B_3 \ldots \quad B_n) \tag{35}$$

$$S = (S_1 \quad S_2 \quad S_3 \ldots \quad S_n) \tag{36}$$

The results of previous lemmas for mean, variance and auto-covariance of individual and aggregated bandwidth process are still valid for the study of bandwidth in multiservice P2P networks.

### 4.4 Synthetic Bandwidth Generation

In this section we provide a method for synthetic generating bandwidth process. It has some applications; for example when we would like to evaluate the proposed model, we use the results of this section to generate the bandwidth process synthetically in order to compare with measured bandwidth. Recall from section 4.2 that the bandwidth process is represented by the stochastic integration. We consider the summation formula instead of integration, because it can be easily implemented by mathematical softwares.

**Lemma5**: The synthetic bandwidth process is generated by:

$$Bw = \sum B(\tau - t')\Delta_s = \frac{a}{\sqrt{1-Y_{i+1}}}\left( S(0)\left(e^{-\lambda t_{i+1}} - e^{-\lambda t_i}\right) + \frac{\sigma}{\sqrt{2\lambda}}\left(\sqrt{1-e^{-2\lambda t_{i+1}}} - \sqrt{1-e^{-2\lambda t_i}}\right) N(0,1) \right)$$

(37)

For the aggregated case the summation of individual processes yields the aggregated bandwidth process.

### 4.5 Estimation of the proposed Model Parameters

The method of Maximum Likehood Estimation (MLE) estimates parameters by finding the values that maximize likehood function. We would like to estimate three parameters $(n\ \gamma\ \sigma)$. The first parameter corresponds to index of the power law distribution, while the last two parameters correspond to OU process parameters. The lemma 6 proposes how one can estimate these parameters. We suppose that the OU process is zero mean.

**Lemma6**: The parameters of the bandwidth process are estimated as:

$$\hat{\gamma} = \frac{-\ln A}{\Delta_k}$$

(38)

$$\hat{\sigma} = \sqrt{\frac{1}{(n+1)} \frac{2x_0^2\left(1-\exp(-2\lambda\Delta_k)\right) + 2\sum_{k=1}^n x_k - 2\exp(-\lambda\Delta_k)\sum_{k=1}^n x_{k-1}^2}{1-\exp(-2\lambda\Delta_k)}}$$

(39)

$$\hat{n} = 1 + n \left[\sum_{j=1}^n \ln \frac{x_j}{a}\right]^{-1}$$

(40)

## 5. Conclusion

We don't have accurate models for P2P bandwidth. In this paper we proposed a new model for P2P bandwidth that is based on OU process. The model was specified mathematically. We also studied the total bandwidth modeling. Some statistical parameters in both the individual and total bandwidth models were derived. We then provided the queuing analysis for the buffer fed by the bandwidth process. The synthetic generations of the model and bandwidth estimation were also investigated in this paper. Possible applications of the proposed model are bandwidth management and capacity planning in P2P networks. Some of the tasks that are scheduled for the future are:

1. It is of interest to evaluate the model with various measured P2P bandwidths
2. How to minimize file download time in P2P networks is another important issue

## Appendix A

## Proof of lemma1

$$E(Bw) = E\left[\gamma\big(B(\tau'-t)*S(t)\big) + \sigma\lim_{n\to\infty}\sum_{p_{i-1},p_i\in\pi_n} B_{p_{i-1}}(W_{p_i} - W_{p_{i-1}})\right] \tag{41}$$

$$E(Bw) = E\left(\gamma\big(B(\tau'-t)*S(t)\big)\right) + E\left(\sigma\lim_{n\to\infty}\sum_{p_{i-1},p_i\in\pi_n} B_{p_{i-1}}(W_{p_i} - W_{p_{i-1}})\right) \tag{42}$$

$$E(Bw) = E\big(\gamma(B+S)\big) + E\left(\sigma\lim_{n\to\infty}\sum_{p_{i-1},p_i\in\pi_n} B_{p_{i-1}}(W_{p_i} - W_{p_{i-1}})\right) \tag{43}$$

$$\lim_{n\to\infty}\sum_{p_{i-1},p_i\in\pi_n} B_{p_{i-1}}(W_{p_i} - W_{p_{i-1}}) = K \tag{44}$$

$$E(Bw) = E\big(\gamma(B+S)\big) + E(\sigma K) \tag{45}$$

$$E(Bw) = \gamma E(B+S) + \sigma E(K) \tag{46}$$

$$E(Bw) = \gamma\big(E(B) + E(S)\big) + \sigma K' \tag{47}$$

$$Var(Bw) = Var\left[\gamma\big(B(\tau'-t)*S(t)\big) + \sigma\lim_{n\to\infty}\sum_{p_{i-1},p_i\in\pi_n} B_{p_{i-1}}(W_{p_i} - W_{p_{i-1}})\right] \tag{48}$$

$$Var(Bw) = Var\left(\gamma\big(B(\tau'-t)*S(t)\big)\right) + Var\left(\sigma\lim_{n\to\infty}\sum_{p_{i-1},p_i\in\pi_n} B_{p_{i-1}}(W_{p_i} - W_{p_{i-1}})\right) \tag{49}$$

$$Var(Bw) = Var(\gamma(B+S)) + Var(\sigma \lim_{n\to\infty} \Sigma_{p_{i-1},p_i \in \pi_n} B_{p_{i-1}}(W_{p_i} - W_{p_{i-1}})) \quad (50)$$

$$Var(Bw) = Var(\gamma(B+S)) + Var(\sigma K) \quad (51)$$

$$Var(Bw) = \gamma Var(B+S) + \sigma Var(K) \quad (52)$$

Since $B$ and $S$ are independent, we have:

$$Var(Bw) = \gamma(Var(B) + Var(S)) + \sigma K' \quad (53)$$

# Appendix B

# Proof of lemma 2

$$ACV(Bw) = ACV[\gamma(B(\tau'-t)*S(t)) + \sigma \lim_{n\to\infty} \Sigma_{p_{i-1},p_i \in \pi_n} B_{p_{i-1}}(W_{p_i} - W_{p_{i-1}})] \quad (54)$$

$$ACV(Bw) = ACV(\gamma(B(\tau'-t)*S(t))) + ACV(\sigma \lim_{n\to\infty} \Sigma_{p_{i-1},p_i \in \pi_n} B_{p_{i-1}}(W_{p_i} - W_{p_{i-1}})) \quad (55)$$

$$ACV(Bw) = ACV(\gamma(B+S)) + ACV(\sigma \lim_{n\to\infty} \Sigma_{p_{i-1},p_i \in \pi_n} B_{p_{i-1}}(W_{p_i} - W_{p_{i-1}})) \quad (56)$$

$$ACV(Bw) = ACV(\gamma(B+S)) + ACV(\sigma K) \quad (57)$$

$$ACV(Bw) = \gamma ACV(B+S) + \sigma ACV(K) \quad (58)$$

$$ACV(Bw) = \gamma(ACV(B) + ACV(S)) + \sigma K' \quad (59)$$

$$ACV(Bw) = \gamma(ACV(B) + ACV(S)) + \sigma K' \quad (60)$$

$$ACV(Bw) = \gamma\left(ak^{2(H-1)} + \frac{a^2}{2\lambda}e^{-\lambda k}\right) + \sigma K' \quad (61)$$

$$\gamma a = C_1, \frac{a^2}{2} = C_2, \sigma K' = C_3 \quad (62)$$

$$ACV(Bw) = C_1 k^{2(H-1)} + C_2 e^{-\lambda k} + C_3 \tag{63}$$

According to [15] the Hurst parameter is given by $H = \dfrac{4 - \min(n_0, n_1)}{2}$ \hfill (64)

so we have: $ACV(Bw) \quad C_1 k^{2\left(\frac{4-\min(n_0,n_1)}{2} - 1\right)} + C_2 e^{-\lambda k} + C_3$ \hfill (65)

We can approximate Eq.65 with:

$$ACV(Bw) \quad C_1 k^{2\left(\frac{4-\min(n_0,n_1)}{2} + \varepsilon - 1\right)} \tag{66}$$

The auto-covariance of the bandwidth process indicates that the $Bw$ is LRD.

## Appendix C

## Proof of lemma3

The mean of aggregated bandwidth process is defined as:

$$E(Bw_t) = E\left(\sum_{i=1}^{N} Bw_i\right) = \sum_{i=1}^{N} E(Bw_i) \tag{67}$$

$$E(Bw_t) = \sum_{i=1}^{N} \left(\gamma_i (E(B_i) + E(S_i)) + \sigma_i K'_i\right) \tag{68}$$

The variance of aggregated bandwidth process is defined as:

$$Var(Bw_t) = Var\left(\sum_{i=1}^{N} Bw_i\right) = \sum_{i=1}^{N} Var(Bw_i) \tag{69}$$

$$Var(Bw_t) = \sum_{i=1}^{N} \gamma (Var(B) + Var(S)) + \sigma K' \tag{70}$$

## Appendix D

## Proof of lemma 4

Suppose we have a queue that is fed by Bw with service rate C, the length of queue is V. We have:

$$V(t) = \sup(A(t) - A(s) - C(t-s)) \tag{71}$$

From [22] we have:

$$P(V(t) \succ x) \quad \exp\left(-\left[\frac{-1}{2ma(1-H)^2}\left(\frac{(1-m)(1-H)}{H}\right)^{2H}\right]\right) x^{2(1-H)} \tag{72}$$

According to lemma1 we have:

$$m = \max\left(\frac{1}{download\,rate}, \frac{1}{upload\,rate}\right)$$

$$H = \frac{4 - \min(n_0, n_1)}{2} + \varepsilon$$

$$a = \gamma(Var(B) + Var(S)) + \sigma K' \tag{73}$$

After substituting these values in Eq.72 we have:

$$P(V(t) \succ x) \quad \exp\left(-\left[\frac{-1}{2m\left(\lambda(Var(B)+Var(B)+\sigma K')\right)\left(\frac{\min(n_0,n_1)}{2}-1-\varepsilon\right)^2}\left(\frac{(1-m)\left(\frac{\min(n_0,n_1)}{2}-1-\varepsilon\right)}{\frac{4-\min(n_0,n_1)}{2}+\varepsilon}\right)^{(4-\min(n_0,n_1)+2\varepsilon)}\right]\right) x^{(\min(n_0,n_1)-2-2\varepsilon)}$$

$$\tag{74}$$

## Appendix E

## Proof of lemma5

$$Bw = \sum_{i=1}^{N} B(\tau - t')\Delta_S = \sum_{i=1}^{N} B(\tau - t')(S_{t_{i+1}} - S_{t_i}) \tag{75}$$

$$\tau - t' = t_{i+1} \tag{76}$$

$$Bw = \sum_{i=1}^{N} B(\tau - t')\Delta_S = \sum_{i=1}^{N} B(t_{i+1})(S_{t_{i+1}} - S_{t_i}) \tag{77}$$

Referring to [15], the $B(t_{i+1})$ is generated by:

$$x_{i+1} = \frac{a}{\sqrt[n]{1-Y_{i+1}}} \tag{78}$$

$Y_{i+1}$ is the random variable with uniform distribution in the range $(0\ \ 1]$ and it can be generated by conventional pseudo-random number generators. Referring to section 3.2, we have:

$$S(t) = S(0)e^{-\gamma t} + \mu(1 - e^{-\gamma t}) + \sigma\sqrt{\frac{1-e^{-2\gamma t}}{2\gamma}} N(0,1) \tag{79}$$

So we have:

$$Bw = \sum B(\tau - t')\Delta_s = \frac{a}{\sqrt{1-Y_{i+1}}}\left(S(0)\left(e^{-\lambda_{i+1}} - e^{-\lambda_i}\right) + \mu\left(e^{-\lambda_i} - e^{-\lambda_{i+1}}\right) + \frac{\sigma}{\sqrt{2\lambda}}\left(\sqrt{1-e^{-2\lambda_{i+1}}} - \sqrt{1-e^{-2\lambda_i}}\right)N(0,1)\right) \tag{80}$$

Since OU process is zero mean, we have:

$$Bw = \sum B(\tau - t')\Delta_s = \frac{a}{\sqrt{1-Y_{i+1}}}\left(S(0)\left(e^{-\lambda_{i+1}} - e^{-\lambda_i}\right) + \frac{\sigma}{\sqrt{2\lambda}}\left(\sqrt{1-e^{-2\lambda_{i+1}}} - \sqrt{1-e^{-2\lambda_i}}\right)N(0,1)\right) \tag{82}$$

## Appendix F

## Proof of lemma 6

We have: $Bw = \gamma(B(\tau' - t) * S(t)) + K$

Since parameters of the proposed bandwidth model are independent, we can separately use MLE method to estimate the parameters. The Maximum likehood function is defined as:

$$\theta_n = \arg\max_\theta L(\theta) = \arg\max_\theta \ln L(\theta) \quad \text{where } L_\theta = \prod_{i=1}^N f(m_i|\theta) \tag{82}$$

We first use MLE to estimate OU process parameters. Then we estimate index of $B$.

Step1: The estimation of OU process:

The likehood function of zero mean OU process is [19]:

$$K(\theta) = -\frac{n+1}{2}\ln(n\sigma^2) - \frac{x_0^2}{\sigma^2} - \frac{1}{2}\sum_{k=1}^{n}(1-\exp(-2\lambda\Delta_k)) - \sum_{k=1}^{n}\frac{x_k - \exp(-\lambda\Delta_k)x_{k-1}^2}{\sigma^2(1-\exp(-2\lambda\Delta_k))} \tag{83}$$

Where $\Delta_k = t_k - t_{k-1}$

$$\frac{\partial k}{\partial \gamma} = 0 \rightarrow$$

$$-\frac{1}{2}\sum_{k=1}^{n} -2\Delta_k(\exp(-2\gamma\Delta_k))$$

$$-\sum_{k=1}^{n}\frac{(x_{k-1}^2\Delta_k\exp(-\gamma\Delta_k))(\sigma^2(1-\exp(-2\gamma\Delta_k))) - (2\Delta_k\exp(-2\gamma\Delta_k))(x_k - \exp(-\gamma\Delta_k)x_{k-1}^2)}{[\sigma^2(1-\exp(-2\gamma\Delta_k))]^2} = 0 \tag{84}$$

$$\sum_{k=1}^{n}\Delta_k(\exp(-2\gamma\Delta_k))[\sigma^2(1-\exp(-2\gamma\Delta_k))]^2 - (x_{k-1}^2\Delta_k\exp(-\gamma\Delta_k))(\sigma^2(1-\exp(-2\gamma\Delta_k))) -$$
$$(2\Delta_k\exp(-2\gamma\Delta_k))(x_k - \exp(-\gamma\Delta_k)x_{k-1}^2) = 0 \tag{85}$$

We assume that all $\Delta_k$ are equal and constant. $\exp(-\gamma\Delta_k) = A \rightarrow \exp(-2\gamma\Delta_k) = A^2$ \hfill (86)

$$\sum_{k=1}^{n}\Delta_k(A^2)[\sigma^2(1-A^2)]^2 - (x_{k-1}^2\Delta_k A)(\sigma^2(1-A^2)) - (2\Delta_k A^2)(x_k - Ax_{k-1}^2) = 0 \tag{87}$$

$$\sum_{k=1}^{n}\Delta_k(A^2)\sigma^4(1 - 2A^2 + A^4) - (x_{k-1}^2\Delta_k A)(\sigma^2(1-A^2)) - (2\Delta_k A^2)(x_k - Ax_{k-1}^2) = 0 \tag{88}$$

$$\sum_{k=1}^{n}\Delta_k(A)\sigma^4(1 - 2A^2 + A^4) - (x_{k-1}^2\Delta_k)(\sigma^2(1-A^2)) - (2\Delta_k A)(x_k - Ax_{k-1}^2) = 0 \tag{89}$$

$$\sum_{k=1}^{n}(\Delta_k\sigma^4 A - 2\Delta_k\sigma^4 A^3 + \Delta_k\sigma^4 A^5) - \sigma^2 x_{k-1}^2\Delta_k - x_{k-1}^2\Delta_k A^2 - (x_k 2\Delta_k A - 2x_{k-1}^2\Delta_k A^2) = 0 \tag{90}$$

$$\sum_{k=1}^{n}(\sigma^4 - x_{k-1}^2)A - 2\sigma^4 A^3 + \sigma^4 A^5 - \sigma^2 x_{k-1}^2 + x_{k-1}^2 A^2 = 0 \tag{91}$$

$$\sum_{k=1}^{n}\sigma^4 A^5 - \sum_{k=1}^{n}2\sigma^4 A^3 + \sum_{k=1}^{n}x_{k-1}^2 A^2 + \sum_{k=1}^{n}(\sigma^4 - x_k^2)A - \sum_{k=1}^{n}\sigma^2 x_{k-1}^2 = 0 \tag{92}$$

$$\sum_{k=1}^{n}\sigma^4 = C_1, \sum_{k=1}^{n}2\sigma^4 = C_2, \sum_{k=1}^{n}x_{k-1}^2 = C_3, \sum_{k=1}^{n}(\sigma^4 - x_k^2) = C_4, \sum_{k=1}^{n}\sigma^2 x_{k-1}^2 = C_5 \tag{93}$$

We also have: $C_2 = 2C_1, C_5 = \sigma^2 C_3$ \hfill (94)

$$C_1 A^5 - C_2 A^3 + C_3 A^2 + C_4 A - C_5 = 0 \tag{95}$$

$$C_1 A^4 - 2C_1 A^2 + C_3 A^1 + C_4 = \frac{\sigma^2 C_3}{A} \tag{96}$$

$$C_1 A^4 - 2C_1 A^2 + C_4 = \frac{\sigma^2 C_3}{A} - C_3 A \tag{97}$$

$$C_1 A^4 - 2C_1 A^2 + C_4 = f_1, \quad \frac{\sigma^2 C_3}{A} - C_3 A = f_2 \tag{98}$$

The solution for A can be easily obtained in mathematical softwares such as Maple by intersection of $f_1$ and $f_2$.

Since $\exp(-\gamma \Delta_k) = A \rightarrow \hat{\gamma} = \frac{-\ln A}{\Delta_k}$ (99)

From the solution of above equation solution since $\hat{\gamma} > 0$, A must be between 0 and 1, that is:
$0 < A < 1$

$$\frac{\partial k}{\partial \sigma} = 0 \rightarrow$$

$$-\frac{n+1}{\sigma} + \frac{2}{\sigma^3} + 2\sum_{k=1}^{n} \frac{x_k - \exp(-\lambda \Delta_k) x_{k-1}^2}{\sigma^3 (1 - \exp(-2\lambda \Delta_k))} = 0 \tag{100}$$

$$-(n+1)\sigma^2 + 2x_0^2 + 2\sum_{k=1}^{n} \frac{x_k - \exp(-\lambda \Delta_k) x_{k-1}^2}{(1 - \exp(-2\lambda \Delta_k))} = 0 \tag{101}$$

$$-(n+1)\sigma^2 + 2x_0^2 + 2\sum_{k=1}^{n} \frac{x_k}{1 - \exp(-2\lambda \Delta_k)} - 2\sum_{k=1}^{n} \frac{\exp(-\lambda \Delta_k) x_{k-1}^2}{1 - \exp(-2\lambda \Delta_k)} = 0 \tag{102}$$

$$-(n+1)\sigma^2 + 2x_0^2 + \frac{2}{1-\exp(-2\lambda \Delta_k)} \sum_{k=1}^{n} x_k - 2\frac{\exp(-\lambda \Delta_k)}{1-\exp(-2\lambda \Delta_k)} \sum_{k=1}^{n} x_{k-1}^2 = 0 \tag{103}$$

$$2x_0^2 + \frac{2}{1-\exp(-2\lambda \Delta_k)} \sum_{k=1}^{n} x_k - 2\frac{\exp(-\lambda \Delta_k)}{1-\exp(-2\lambda \Delta_k)} \sum_{k=1}^{n} x_{k-1}^2 = (n+1)\sigma^2 \tag{104}$$

$$\frac{2x_0^2 (1-\exp(-2\lambda \Delta_k)) + 2\sum_{k=1}^{n} x_k - 2\exp(-\lambda \Delta_k) \sum_{k=1}^{n} x_{k-1}^2}{1-\exp(-2\lambda \Delta_k)} = (n+1)\sigma^2 \tag{105}$$

Recall that $\exp(-\gamma \Delta_k) = A \rightarrow \exp(-2\gamma \Delta_k) = A^2$ (106)

$$\frac{2x_0^2 (1-A^2) + 2\sum_{k=1}^{n} x_k - 2A \sum_{k=1}^{n} x_{k-1}^2}{1-A^2} = (n+1)\sigma^2 \tag{107}$$

$$\frac{-2x_0^2 A^2 - 2A \sum_{k=1}^{n} x_{k-1}^2 + 2 \sum_{k=1}^{n} x_k + 2x_0^2}{1 - A^2} = (n+1)\sigma^2 \qquad (108)$$

$$\frac{x_0^2 A^2 + A \sum_{k=1}^{n} x_{k-1}^2 - \sum_{k=1}^{n} x_k - x_0^2}{A^2 - 1} = \frac{(n+1)}{2}\sigma^2 \qquad (109)$$

$$\Delta = \left( \sum_{k=1}^{n} x_{k-1}^2 \right)^2 + 4x_0^2 \left( \sum_{k=1}^{n} x_k + x_0^2 \right) \qquad (110)$$

$$A = \frac{-\sum_{k=1}^{n} x_{k-1}^2 \pm \sqrt{\Delta}}{2x_0^2} \qquad (111)$$

$$\text{If } A \succ \frac{-\sum_{k=1}^{n} x_{k-1}^2 + \sqrt{\left( \sum_{k=1}^{n} x_{k-1}^2 \right)^2 + 4x_0^2 \left( \sum_{k=1}^{n} x_k + x_0^2 \right)}}{2x_0^2} \qquad (112)$$

then $\sigma$ has a solution. Only the positive solution is considered. We have:

$$\frac{1}{(n+1)} \frac{2x_0^2 \left(1 - \exp(-2\lambda \Delta_k)\right) + 2\sum_{k=1}^{n} x_k - 2\exp(-\lambda \Delta_k) \sum_{k=1}^{n} x_{k-1}^2}{1 - \exp(-2\lambda \Delta_k)} = \sigma^2 \qquad (113)$$

$$\hat{\sigma} = \sqrt{\frac{1}{(n+1)} \frac{2x_0^2 \left(1 - \exp(-2\lambda \Delta_k)\right) + 2\sum_{k=1}^{n} x_k - 2\exp(-\lambda \Delta_k) \sum_{k=1}^{n} x_{k-1}^2}{1 - \exp(-2\lambda \Delta_k)}} \qquad (114)$$

Step 2: The index estimation of power law distribution

The distribution we used in [15] for modeling P2P traffic is: $f(x) = \begin{cases} 0 & x \prec a \\ k/x^n & x \geq a \end{cases}$ (115)

The maximum likehood estimation for distribution like this is represented by [23]:

$$\hat{n} = 1 + n \left[ \sum_{j=1}^{n} \ln \frac{x_j}{a} \right]^{-1}$$ (116)

and the proof is completed.